\documentclass[12pt]{article}
\pdfoutput=1
\usepackage{geometry,enumerate,amsmath,amssymb}
\usepackage{fullpage}
\usepackage{graphicx}
\usepackage{bm}
\usepackage[colorlinks=false, urlcolor=blue, linkcolor=red, hidelinks]{hyperref}
\numberwithin{equation}{section}
\newcommand{\be}{\begin{equation}}
\newcommand{\ee}{\end{equation}}
\newcommand{\bea}{\begin{eqnarray}}
\newcommand{\eea}{\end{eqnarray}}
\renewcommand{\epsilon}{\varepsilon}

\newcommand{\bsigma}{\boldsymbol{\sigma}}
\newcommand{\bn}{\boldsymbol{n}}

\begin{document}
\title{Monopole fission}
\author{
  Paul Sutcliffe\\[10pt]
  {\em \normalsize Department of Mathematical Sciences,}\\
{\em \normalsize Durham University, Durham DH1 3LE, United Kingdom.}\\
 {\normalsize Email:  p.m.sutcliffe@durham.ac.uk}
}

\date{September 2026}

\maketitle
\begin{abstract}
 There is a \(4N\)-dimensional moduli space of \(SU(2)\) BPS monopoles with charge \(N\), but introducing a Higgs potential removes this degeneracy and yields a repulsive force between monopoles. However, if a point in the BPS monopole moduli space is sufficiently symmetric, then it flows to a similar non-BPS saddle point solution once the Higgs potential is introduced. Monopole fission, into constituent monopoles with lower charges, results from 
 symmetry breaking perturbations of such saddle point solutions. This fission is investigated numerically, by evolving the gradient flow equations of the Yang-Mills-Higgs theory, with initial conditions and symmetry breaking perturbations created using rational maps between Riemann spheres. The results resemble components of BPS monopole scattering processes obtained within the geodesic approximation for low speed monopole dynamics.

\end{abstract}

\newpage

\section{Introduction}\quad
In the BPS limit of a vanishing Higgs potential there is a moduli space \(M_N\), with dimension \(4N\), of minimal energy \(SU(2)\) monopoles with charge \(N\). A point in  \(M_N\) corresponds to a solution of the first-order Bogomolny equation, and the \(4N\) parameters may be interpreted as a position and a \(U(1)\) phase for each of the \(N\) unit charge monopoles, in the regime in which these are all well-separated. There are points in \(M_N\) where the monopoles merge to produce a solution with enhanced symmetry, including an axially symmetric monopole for all \(N\ge 2\) and various platonic monopoles for particular values of \(N\). For a review see \cite{book}.

The introduction of a Higgs potential no longer allows static solutions of the second-order Yang-Mills-Higgs equations to be obtained as solutions of a first-order Bogomolny equation. The large moduli space of energy degenerate solutions is lost because the Higgs potential results in a repulsive force between monopoles, so it is reasonable to expect that there are no stable solutions that describe merged monopoles. However, if a point in \(M_N\) has sufficient symmetry, then it will flow to a saddle point solution with the same symmetry upon the introduction of the Higgs potential. This will be the case if the symmetry is incompatible with monopole fission into any constituent monopoles with lower charges. 

These saddle point solutions can be computed numerically by evolving the gradient flow equations of the Yang-Mills-Higgs theory from suitable initial conditions. Furthermore, symmetric monopole fission is realized by applying controlled symmetry breaking perturbations. This is achieved by using rational maps between Riemann spheres to provide monopole fields for the initial conditions of the gradient flow. An indication of the accuracy of these rational map generated approximate initial monopoles is obtained by application to the BPS limit, where it is found that the energy is typically of the order of around \(15\%\) above the Bogomolny bound, although the energy excess can be below \(10\%\). As expected, given the decent approximation of the initial conditions, the gradient flow quickly yields the corresponding symmetric BPS monopole. This approach is therefore also an efficient method to numerically compute a variety of BPS monopoles.

Section 2 describes the rational map approximation to monopoles and discusses some of its features when applied to the case of BPS monopoles. Section 3 presents 
the numerical scheme for solving the Yang-Mills-Higgs gradient flow equations, and the results that it produces on symmetric monopole fission for a number of examples, mainly considering saddle point solutions with platonic symmetry and their fission under perturbations that break the platonic symmetry to a variety of subgroups. 
These results reveal a qualitative agreement with components of BPS monopole scattering processes obtained within the geodesic approximation \cite{Ma1} for low speed monopole dynamics. Finally, Section 4 contains some concluding remarks.

\section{Monopoles and rational maps}\quad
The static energy of the \(SU(2)\) Yang-Mills-Higgs theory is given by
\be
E=\int \bigg(
 -\frac{1}{2}\mbox{Tr}(D_i\Phi D_i\Phi)
 -\frac{1}{4}\mbox{Tr}(F_{ij}F_{ij})
 +\frac{\lambda}{2}(1-|\Phi|^2)^2
 \bigg) d^3x,
 \label{energy}
\ee
where \(\Phi,A_i,\) are the \(\mathfrak{su}(2)\)-valued Higgs field and gauge potential, with \(D_i\Phi=\partial_i\Phi+[A_i,\Phi]\) the covariant derivative of the Higgs field, and \(F_{ij}=\partial_i A_j-\partial_j A_i+[A_i,A_j]\) the field strength.

For fields with finite energy, the Higgs field on the sphere at infinity is a map between two-spheres, with the winding number, \(N\in \pi_2(S^2)=\mathbb{Z},\) referred to as the monopole charge. Without loss of generality, the restriction to positive charge will be made in this study. Completing the square with the first two terms in the energy density yields the Bogomolny bound 
\(E\ge 4\pi N,\) but this cannot be attained unless the non-negative parameter \(\lambda\) in the Higgs potential vanishes, which is known as the BPS limit.

The static field equations that follow from the variation of the energy (\ref{energy}) are
\be
D_iD_i\Phi=-\lambda \Phi(1-|\Phi|^2), \qquad\qquad
D_j F_{ji}=[\Phi,D_i\Phi].
\label{static}
\ee
In the BPS limit, solutions of the second-order field equations (\ref{static}) can be obtained by solving the first-order Bogomolny equation
\be
D_i\Phi=\frac{1}{2}\varepsilon_{ijk}F_{jk},
\label{Bog}
\ee
producing charge \(N\) monopoles that attain the energy bound, \(E=4\pi N.\) The \(4N\)-dimensional moduli space, \(M_N,\) of these BPS monopoles includes \(N\) unit charge monopoles with arbitrary positions and phases, reflecting the fact that there are no static forces between BPS monopoles. Although the overall phase can be changed by a global action of the unbroken \(U(1)\) gauge group, it is convenient to keep this as one of the \(4N\) moduli.  

There is a description of the moduli space \(M_N\) in terms of degree \(N\) rational maps between Riemann spheres \cite{Jar}, that respects the action of spatial rotations around the origin in \(\mathbb{R}^3\). Briefly, the rational map appears as scattering data along radial half-lines from the origin, as follows. Given the fields of a charge \(N\) BPS monopole, use spherical polar coordinates in \(\mathbb{R}^3\) given by the radius \(r\) and the Riemann sphere coordinate \(z\). Along radial half-lines, given by fixing the value of \(z\), consider the solution of the scattering equation
\be
(D_r-i\Phi)\begin{pmatrix} w_1 \\ w_2 \end{pmatrix}=0,
\label{scat}
\ee
that decays as \(r\to\infty,\) which is unique up to the multiplication by an overall factor that is independent of \(r.\) It can be shown that evaluating the ratio of the two components at the origin gives a degree \(N\) rational map in \(z\), that is,
\(R(z)=({w_1}/{w_2})\big|_{r=0}.\) A gauge transformation, \(g\in SU(2)\), of the monopole fields acts on the rational map as an \(SU(2)\) M\"obius transformation
\be
R(z)\mapsto \widetilde R(z)=\frac{\alpha R(z)+\beta}{-\bar\beta R(z)+\bar\alpha}, \quad \mbox{with} \ \
g_0=\begin{pmatrix} \alpha & \beta \\ -\bar\beta & \bar\alpha
\end{pmatrix}\in SU(2),
\label{mobius}
\ee
where \(g_0\) is the gauge transformation evaluated at the origin. Two rational maps are therefore regarded as equivalent if they are related by an \(SU(2)\) M\"obius transformation, and the equivalence class of degree \(N\) rational maps has dimension \(4N-1.\) Adding back in the overall \(U(1)\) phase, that is included within the definition of the moduli space \(M_N\), recovers the correct \(4N\) dimensions, giving a one-to-one correspondence between BPS monopoles and rational maps. 

As the rational map of a monopole arises as scattering data there is no direct expression for the monopole fields in terms of the rational map. However, there is an explicit approximation for the monopole fields using the rational map, that appears to be a reasonable description in the situation where all the monopoles have merged at the origin. To present this approximation, let \(\bn\) be the unit vector obtained by inverse stereographic projection of the Riemann sphere coordinate \(R(z),\) 
\be
\bn=\frac{1}{1+|R|^2}(R+\bar R,i(R-\bar R),|R|^2-1).
\ee
The approximation for the monopole fields is 
\be
\Phi= ih\,\bn\cdot\bsigma, \qquad
A_i=\frac{i}{2}(1-k)(\bn\times \partial_i \bn)\cdot\bsigma,
\label{approx}
\ee
where \(\bsigma\) denotes the triple of Pauli matrices, with \(h(r)\) and \(k(r)\) real radial profile functions satisfying the boundary conditions \(h(0)=0,\,h(\infty)=1,\,k(0)=1,\,k(\infty)=0.\) Note that the winding number of the Higgs field on the sphere at infinity is equal to the degree \(N\) of the rational map \(R(z)\), hence this approximate monopole has charge \(N.\)
In the case \(N=1\), and taking \(R(z)=z\) for the rational map, (\ref{approx}) simplifies to the usual hedgehog ansatz, so the exact charge one monopole solution is recovered for appropriate profile functions, in both the BPS and non-BPS situations.

For \(N>1\) there are no solutions of the static field equations (\ref{static}) of the form (\ref{approx}), but substituting this approximation into the energy (\ref{energy}) and performing the angular integration gives 
\be
E=\pi\int \bigg(
4h'^2r^2+2N(4h^2k^2+k'^2)+\frac{{\cal I}}{r^2}(1-k^2)^2
+2\lambda (1-h^2)^2r^2
\bigg)dr,
\label{renergy}
\ee
where prime denotes differentiation with respect to \(r\), and \({\cal I}\) is the integral
\be
{\cal I}=\frac{1}{4\pi}\int \bigg(
\frac{(1+|z|^2)}{(1+|R|^2)}\bigg|\frac{dR}{dz}\bigg|
\bigg)^4\frac{2idzd\bar z}{(1+|z|^2)^2},
\ee
which defines an energy function on the space of degree \(N\) rational maps. 

The same quantity \({\cal I}\) appears in the energy expression for a rational map approximation to Skyrmions \cite{HMS}, so its properties have been studied extensively in that context \cite{BS3}. Some results that are relevant to the present study are presented in Table 1, for \(1\le N\le 7.\) This includes the minimal value of \({\cal I}/N^2\), which is bounded below by 1, and the symmetry group \(G\) of the \({\cal I}\) minimizing rational map. Note that a rational map \(R(z)\) is symmetric under a rotation, which acts as an 
\(SU(2)\) M\"obius transformation \(z\mapsto \widetilde z\), if this produces a map equivalent to \(R(z)\), namely,
\(R(\widetilde z)=\widetilde R(z),\) for some 
compensating \(SU(2)\) M\"obius transformation (\ref{mobius}), which need not be the same transformation that acts on \(z.\) The required minimal maps can be found in Section 3, where their symmetries are discussed in detail, together with some symmetry breaking perturbations.

\begin{table}[h!]
\begin{center}
  \begin{tabular}{|c|c|c|c|}
    \hline
    \(N\) & \(G\) & \({\cal I}/N^2\) & \(E/(4\pi N)\) \\
    \hline
    1 & \(O(3)\) & 1.00 & 1.00\\
  2 & \(D_{\infty h}\) & 1.45 & 1.14\\
   3 & \(T_d\) & 1.51 & 1.17\\
    4 & \(O_h\) &  1.29 & 1.10\\
     5 & \(D_{2d}\) & 1.43 & 1.15\\
      6 &  \(D_{4d}\) & 1.41 & 1.15\\
       7 & \(Y_h\) &  1.24&1.09 \\
    \hline
  \end{tabular}
\caption{For each charge  \(1\le N\le 7\), the symmetry group \(G\) of the \({\cal I}\) minimizing rational map, the value of \({\cal I}/N^2\), and the ratio of the energy to the Bogomolny bound in the BPS limit for the monopole fields generated using the rational map approximation (\ref{approx}).}
\end{center}
\end{table}

For any given value of \(N\), the energy (\ref{renergy}) is minimized by using the rational map that minimizes \({\cal I}\), so in that sense the minimizing map provides the best approximation of the form (\ref{approx}) to a charge \(N\) monopole. Using the value of \({\cal I}\), the profile functions, \(h(r),k(r),\) satisfying the boundary conditions given earlier, are computed numerically using a standard gradient flow method, to obtain the approximate monopole and its energy. The results of this procedure will be applied in Section 3 in the non-BPS case \(\lambda>0,\) to provide initial conditions for fully three-dimensional simulations. However, it is perhaps useful to obtain an indication of the accuracy of this approximation by first applying it to the BPS case \(\lambda=0,\) where the exact value of the monopole energy is known, \(E=4\pi N,\) because the Bogomolny bound is attained. The results are presented in the final column of Table 1, where the ratio of the energy of the approximation to the exact value is shown. This reveals that the energy of the approximation is typically around \(15\%\) above the true value for \(N>1\), although it can be below \(10\%.\) Naturally, the energy excess correlates with the excess of \({\cal I}/N^2\) over the bound of unity. The expectation is that the errors are similar in the non-BPS case. 

It is perhaps worthwhile making a comment on the difference between the rational map approximation for Skyrmions and for BPS monopoles. Skyrmions are attractive (for suitable relative orientations) and form bound states in which the individual Skyrmions merge. The \({\cal I}\) minimizing rational maps, with suitable profile functions, provide good approximations to these minimal energy Skyrmions, typically with an error of around a few percent. Furthermore, the symmetries of the minimizing rational maps matches the symmetries of the true Skyrmion solutions obtained via numerical simulations \cite{BS3}. There is therefore a correlation between the energy function \({\cal I}\) on the space of rational maps and the energy of Skyrmions.

In the case of BPS monopoles, all points in the moduli space \(M_N\) have the same energy, so the interpretation of \({\cal I}\) as an energy function on the space of rational maps does not reflect an energy function on the space of BPS monopoles. Rather, it provides an indication of how close the rational map approximate monopole fields are to a point in \(M_N.\) The expectation is that the 
\({\cal I}\) minimizing map provides the monopole fields for the best approximation to a point in \(M_N.\) Rational maps with a value of \({\cal I}\) significantly above the 
minimal value are not expected to provide monopole fields that are close to any point in \(M_N\), when used with the approximation (\ref{approx}).
Given that there is a correspondence between \(M_N\) and rational maps via scattering data, a reasonable expectation is that if the rational map approximation (\ref{approx}) does provide fields close to some point in \(M_N\), then it is likely to be the point with the given rational map as its scattering data. As supporting evidence for this expectation, the scattering data of the approximate monopole fields (\ref{approx}) is derived below.

First note that the fields (\ref{approx}) are in radial gauge, 
\be
A_r=\frac{x_iA_i}{r}=\frac{i}{2r}(1-k)(\bn\times x_i\partial_i \bn)\cdot\bsigma
=\frac{i}{2}(1-k)(\bn\times \partial_r \bn)\cdot\bsigma
=0,
\ee
as \(\bn\) is independent of \(r.\) Therefore, substituting
the approximate fields (\ref{approx}) into the scattering equation (\ref{scat}), considered along the radial half-line with \(z\) constant, gives 
\be
\bigg(\partial_r +\frac{h(r)}{1+|R|^2}
\begin{pmatrix}|R|^2-1 & 2R \\ 2\bar R & 1-|R|^2\end{pmatrix}
\bigg)
\begin{pmatrix}w_1 \\ w_2 \end{pmatrix}=0.
\ee
The solution that decays as \(r\to\infty\) is
\be
\begin{pmatrix}w_1 \\ w_2 \end{pmatrix}
=
\begin{pmatrix}R \\ 1 \end{pmatrix}
\chi(z,\bar z)\exp{\bigg(-\int_0^r h(r')\,dr'\bigg)},
\ee
where \(\chi(z,\bar z)\) is an arbitrary function of \(z\) and \(\bar z.\) Hence the rational map scattering data
is \((w_1/w_2)|_{r=0}=R(z),\) and is indeed equal to the rational map used in the approximation.

\section{Fission from symmetry breaking perturbations}\quad
The previous section described the construction of initial conditions, from rational maps, that will be used in this section as starting points for the evolution of the gradient flow equations that follow from the energy (\ref{energy}), namely
\bea
&&\frac{\partial \Phi}{\partial t}=D_iD_i\Phi+\lambda \Phi(1-|\Phi|^2),\label{gradflow1}\\
&&\frac{\partial A_i}{\partial t}=
D_j F_{ji}+[D_i\Phi,\Phi]+\partial_i\partial_j A_j.
\label{gradflow2}
\eea
The last term in (\ref{gradflow2}) is a gauge fixing term that ensures that the flow equations are parabolic.
The initial conditions (\ref{approx}) are already in Coulomb gauge, as the following short calculation confirms,
\bea
\partial_j A_j&=&-\frac{i}{2r}k'(\bn\times x_j\partial_j \bn)\cdot\bsigma+\frac{i}{2}(1-k)(\bn\times \partial_j\partial_j \bn)\cdot\bsigma\nonumber\\
&=&\frac{i}{2r^2}(1-k)(1+|z|^2)^2(\bn\times \partial_z\partial_{\bar z} \bn)\cdot\bsigma=0,
\eea
using the fact that \(\bn\) is independent of \(r\) and
\((1+|R|^2)^2\partial_z\partial_{\bar z} \bn=-2\bn|dR/dz|^2.\)

The equations (\ref{gradflow1}) and (\ref{gradflow2}) are solved numerically on a cubic grid containing \(101^3\) lattice points with a lattice spacing \(\Delta x=0.2\). Spatial derivatives are computed using second-order accurate finite difference approximations, with a 27-point stencil for the Laplacian, and time evolution is via a first-order Euler method with a timestep \(\Delta t=(\Delta x)^2/10.\) Homogeneous Neumann boundary conditions are imposed on all fields at the boundary of the grid.

As a check on the numerical method, in the BPS limit (\(\lambda=0\)) it is found that using the \({\cal I}\) minimizing maps associated with the values in Table 1, the flow rapidly converges within a time of order one to the corresponding BPS monopole solution of the static equations (\ref{static}). In particular, energy density isosurface plots are in good agreement with those obtained for BPS monopoles using other methods, such as a numerical Nahm transform \cite{HS1}. This approach appears to be an efficient method to numerically compute a variety of BPS monopoles.

Turning now to the non-BPS case, the results will be presented for the value \(\lambda=1,\) although other values have also been investigated and produce the same qualitative features. Initial conditions using the \({\cal I}\) minimizing maps associated with Table 1 yield saddle point solutions for \(N>1\) with the same symmetry as the rational map, and similar qualitative features to the corresponding BPS monopoles, except for the charges \(N=5\) and \(N=6.\) As expected, for these two charges the dihedral symmetry is insufficient to prevent the flow from separating the constituent monopoles, as a consequence of the repulsive force between monopoles. As there are no saddle point solutions associated with these two rational maps, no further investigations will be reported for monopoles with charges five or six. For the other charges, the results are discussed below and the output of the flow is presented as energy density isosurface plots at increasing times.

\subsection{Axial charge \(N\)}
For \(N=2\) the minimizing rational map has axial symmetry, and the gradient flow relaxes to an axially symmetric saddle point solution. In fact, this is the first member of a family of axially symmetric saddle point solutions for all charges \(N\ge 2,\) which may be considered simultaneously. For any \(N\ge 2\) the rational map \(R(z)=z^N\) is axially symmetric, or more precisely it has \(D_{\infty h}\) symmetry generated by the \(U(1)\) rotation \(R(e^{i\theta}z)=e^{iN\theta}R(z)\), the \(C_2\) rotation \(R(1/z)=1/R(z)\), and the reflection \(R(\bar z)=\overline{R(z)}.\) Relaxing the initial fields generated by this rational map produces an axially symmetric charge \(N\) monopole.

\begin{figure}[!ht]\begin{center}
    \includegraphics[width=0.9\columnwidth]{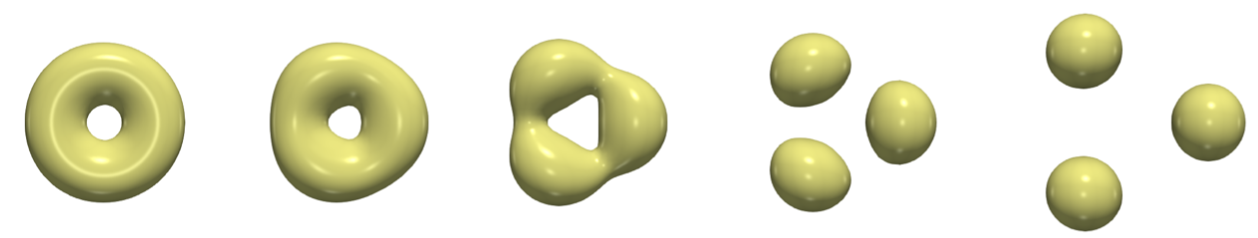}
    \caption{Energy density isosurfaces at increasing times for the axially symmetric charge three monopole with a \(D_{3h}\) symmetric perturbation given by (\ref{pert3axial}) with \(\varepsilon=0.001.\)}
    \label{fig:3d3ha}\end{center}\end{figure}

The most symmetric unstable mode of this saddle point solution is obtained by breaking the \(D_{\infty h}\) symmetry to the dihedral symmetry \(D_{Nh}\), by breaking the \(U(1)\) symmetry to 
\(R(e^{2\pi i/N}z)=R(z)\) via the perturbed rational map
\be
R(z)=\frac{z^N-\varepsilon}{1-\varepsilon z^N},
\label{pert3axial}
\ee
with \(\varepsilon\) real. If \(\varepsilon\) is sufficiently small, then there is an initial transient phase of the flow where the relaxation produces a configuration extremely close to the axially symmetric saddle point solution, followed by a slower evolution realizing monopole fission seeded by the perturbation, and preserving the symmetry of the perturbed map.  The \(N=3\) example is presented in Figure \ref{fig:3d3ha}, using the value \(\varepsilon=0.001\) for the perturbation parameter.
The first energy density isosurface in Figure \ref{fig:3d3ha} shows a configuration very close to the axially symmetric charge three monopole, with the subsequent images revealing the fission into three separate unit charge monopoles on the vertices of an expanding triangle.

There is a geodesic in \(M_N\) with \(D_{Nh}\) symmetry  that describes a related planar scattering of \(N\) monopoles through the axially symmetric charge \(N\) monopole \cite{HMM}, within the geodesic approximation 
for low speed dynamics of BPS monopoles \cite{Ma1}. 
The monopoles approach on the vertices of a contracting regular \(N\)-gon, instantaneously form the axially symmetric monopole, and recede on the vertices of an expanding \(N\)-gon that is rotated through an angle \(\pi/N\) relative to the incoming \(N\)-gon. This is a generalization of the famous right-angle scattering of two monopoles in a head-on collision \cite{AH}.

Gradient flow of the non-BPS axial charge \(N\) monopole, perturbed as above, produces a similar family of monopole configurations. More specifically, it mirrors the component of the geodesic after the formation of the axial monopole. The remaining component is obtained, in reverse, by applying the same perturbation with \(\varepsilon<0.\) It has been verified that setting the parameter \(\lambda\) to zero, after a suitable intermediate time during the flow, yields the numerical computation of the BPS monopole for the associated point on the geodesic, with the \(\lambda=0\) flow projecting the configuration to a point in the moduli space \(M_N\). This provides a method to compute a wide range of BPS monopoles, beyond those that can be obtained by using the minimizing rational maps and restricting the flow to \(\lambda=0\) from the outset. For example, it can be applied to compute BPS monopoles on the vertices of a regular \(N\)-gon with a prescribed circumradius. 

An application would be to provide initial conditions for simulations of slow motion monopole scattering in the relativistic theory, by computing configurations for the first two timesteps using two slightly different values of the circumradius to specify the initial speed. Simulations of such planar monopole scatterings have recently been reported \cite{Bach} using an alternative approach to generate the initial conditions, via an approximation involving sets of angles and radii based on the set of positions of the individual monopoles. That approximation for the initial cyclic arrangement has the monopole phases fixed so that only planar scatterings are possible from the initial conditions. As described below, the gradient flow approach using perturbed rational maps is applicable to a wide range of scenarios, so it could be applied to generate a variety of initial conditions, including cyclic monopoles that yield non-planar scatterings.

\subsection{Tetrahedral charge 3}
To go beyond planar monopole fission, saddle points with platonic symmetry will be considered, with a variety of perturbations investigated that break the platonic symmetry to some of its subgroups. As a first example,
consider the tetrahedrally symmetric \({\cal I}\) minimizing rational map of degree three,
\be
R(z)=\frac{\sqrt{3}iz^2-1}{z^3-\sqrt{3}iz}.
\ee
The platonic \(T_d\) symmetry of this rational map is generated by the \(C_3\) rotation that acts as  
\(R((iz+1)/(-iz+1))=(iR(z)+1)/(-iR(z)+1)\), the \(C_2\) rotation \(R(-z)=-R(z),\) and the reflection \(R(i\bar z)=i\overline{R(z)}.\) 

The \(T_d\) symmetry is broken to cyclic \(C_{3v}\) symmetry by removing the \(C_2\) generator and using the perturbed rational map
\be
R(z)=\frac{-(1+i)\varepsilon z^3+i\sqrt{3}(1+2\varepsilon)z^2+\sqrt{3}(i-1)\varepsilon z-1-2\varepsilon}{z^3-\sqrt{3}(1+i)\varepsilon z^2-i\sqrt{3}z+(1-i)\varepsilon}.
\label{map3c3}
\ee
\begin{figure}[!ht]\begin{center}
    \includegraphics[width=1.0\columnwidth]{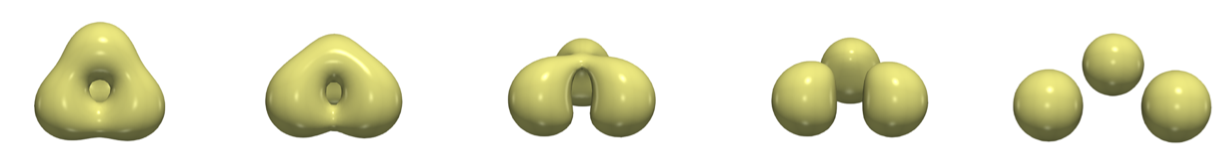}
    \caption{Energy density isosurfaces at increasing times for the tetrahedrally symmetric charge three monopole with a \(C_{3v}\) symmetric perturbation given by (\ref{map3c3}) with \(\varepsilon=0.01.\)}
    \label{fig:3c3a}\end{center}\end{figure} 
The resulting gradient flow evolution for \(\varepsilon=0.01\) is presented in Figure \ref{fig:3c3a} as energy density isosurfaces at increasing times. Once again, the first image represents the end of the initial transient flow, displaying a configuration very close to the tetrahedrally symmetric saddle point solution. The perturbation seeds a fission into three single monopoles, again on the vertices of an expanding triangle, but the relative phases between the monopoles are different from those in the triangular arrangement in the fission in Figure \ref{fig:3d3ha}.
\begin{figure}[!ht]\begin{center}
    \includegraphics[width=1.0\columnwidth]{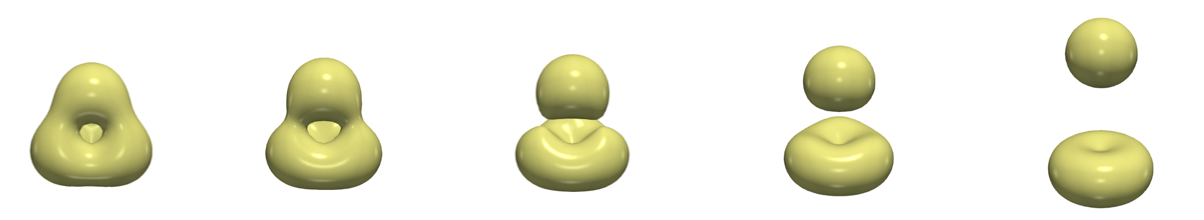}
    \caption{Energy density isosurfaces at increasing times for the tetrahedrally symmetric charge three monopole with a \(C_{3v}\) symmetric perturbation given by (\ref{map3c3}) with \(\varepsilon=-0.01.\)}
    \label{fig:3c3b}\end{center}\end{figure} 
Applying the same perturbation, after changing the sign of the perturbation parameter, produces the evolution displayed in Figure \ref{fig:3c3b}. In this case the fission of the tetrahedral monopole is into a single monopole and a charge two monopole, which move in opposite directions along an axis perpendicular to the plane containing the triangular arrangement of the previous fission. 

Reversing the sequence in Figure \ref{fig:3c3a} and following it by the sequence in Figure  \ref{fig:3c3b} provides a series of configurations similar to those of a geodesic in \(M_3\) describing the \(C_{3v}\) symmetric scattering of three BPS monopoles \cite{HMM}. The Nahm data for this family of BPS monopoles is not known, but approximate Nahm data has been obtained and used to compute energy density isosurfaces \cite{Su5} in qualitative agreement with those in Figures 
\ref{fig:3c3a} and \ref{fig:3c3b}. As mentioned earlier, the initial conditions used in recent numerical simulations \cite{Bach} of monopole scattering, with \(N\) monopoles on the vertices of an \(N\)-gon, only allowed planar scatterings to be investigated. However, that study introduced a second type of initial condition, in which a pair of monopoles approach along a line, with each monopole being an approximation to an axially symmetric monopole of any charge (or a unit charge monopole) with the symmetry axis equal to the line of approach. By using a starting configuration resembling the final image in Figure \ref{fig:3c3b}, the simulations were able to produce the scattering through the tetrahedron to the triangular separation of three single monopoles, with associated energy density isosurfaces similar to those in Figures \ref{fig:3c3a} and \ref{fig:3c3b}.

An alternative fission of the tetrahedral charge three monopole is obtained by breaking the \(T_d\) symmetry to a \(D_{2d}\) subgroup by breaking the \(C_3\) rotation, while preserving the \(C_2\) rotations \(R(-z)=-R(z)\) and \(R(-1/z)=-1/R(z),\) plus the reflection \(R(i\bar z)=i\overline{R(z)}.\) The required perturbed rational map, with \(\varepsilon\) the real perturbation parameter, is
\be
R(z)=\frac{\sqrt{3}i(1+\varepsilon)z^2-1}{z^3-\sqrt{3}i(1+\varepsilon)z}.
\label{map3d2d}
\ee
 \begin{figure}[!ht]\begin{center}
    \includegraphics[width=1.0\columnwidth]{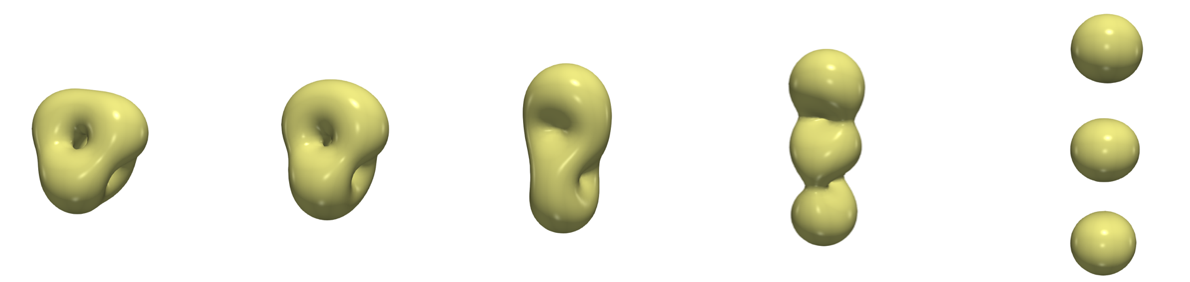}
    \caption{Energy density isosurfaces at increasing times for the tetrahedrally symmetric charge three monopole with a \(D_{2d}\) symmetric perturbation given by (\ref{map3d2d}) with \(\varepsilon=0.1.\)}
    \label{fig:3d2da}\end{center}\end{figure} 

The resulting gradient flow evolution for \(\varepsilon=0.1\) is presented in Figure \ref{fig:3d2da}, revealing the fission to a linear arrangement of three single monopoles. The same \(D_{2d}\) symmetric perturbation (\ref{map3d2d}) with \(\varepsilon<0\) does not yield monopole fission, as it flows to the axially symmetric charge three monopole. This is not surprising as the map (\ref{map3d2d}) with \(\varepsilon=-1\) is the axially symmetric map \(R(z)=-1/z^3\), so this flow is essentially the flow from \(\varepsilon=0\) to \(\varepsilon=-1,\) between two different saddle point solutions.  
 The reverse of the flow in Figure \ref{fig:3d2da}, followed by the flow between the tetrahedron and the axially symmetric monopole, resembles half a geodesic in \(M_3\) describing the twisted line scattering of three BPS monopoles \cite{HS3}, with qualitative agreement with the energy density isosurfaces obtained from the Nahm data of this family of BPS monopoles. The second half of the geodesic is a reversal of the first half with a \(\pi/2\) rotation around the line of approach, so that the dual tetrahedron is formed before the three monopoles separate along the same line of approach.
  
 \subsection{Cubic charge 4}   
 The \({\cal I}\) minimizing map of degree four has cubic symmetry and is given by 
 \be
R(z)=\frac{z^4+2i\sqrt{3}z^2+1}{z^4-2i\sqrt{3}z^2+1},
\label{map4}
 \ee
with the \(O_h\) symmetry generated by the 
\(C_4\) rotation \(R(iz)=1/R(z)\), and the \(C_3\) rotation
\(R((iz+1)/(-iz+1))=e^{2\pi i/3}R(z)\), plus the reflection
\(R(i\bar z)={\overline{R(z)}}.\)

The cubic \(O_h\) symmetry is broken to tetrahedral \(T_d\) symmetry by breaking the \(C_4\) rotation to the \(C_2\) rotation \(R(-z)=R(z),\) using the perturbed rational map 
\be
R(z)=\frac{(1+\varepsilon)(z^4+2i\sqrt{3}z^2+1)}{(1-\varepsilon)(z^4-2i\sqrt{3}z^2+1)}.
\label{map4t}
 \ee
\begin{figure}[!ht]\begin{center}
    \includegraphics[width=1.0\columnwidth]{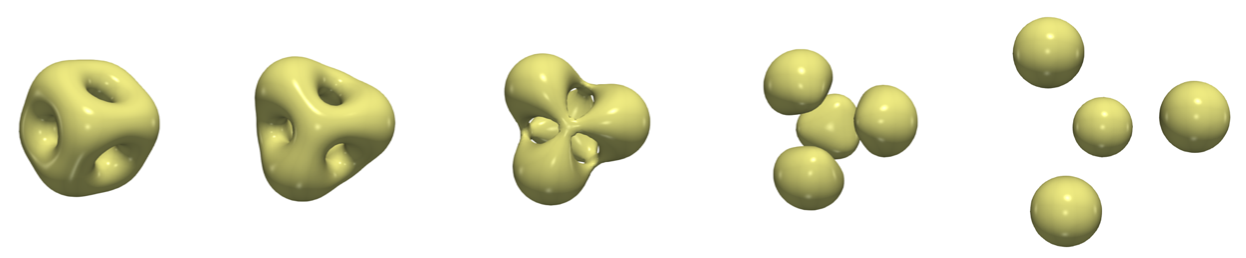}
    \caption{Energy density isosurfaces at increasing times for the cubic charge four monopole with a \(T_{d}\) symmetric perturbation given by (\ref{map4t}) with \(\varepsilon=0.01.\)}
    \label{fig:4t}\end{center}\end{figure} 
    
Figure \ref{fig:4t} displays the gradient flow evolution for \(\varepsilon=0.01,\) with the first image close to the cubic saddle point, which fissions to four single monopoles on the vertices of an expanding tetrahedron. Changing the sign of \(\varepsilon\) is equivalent to the \(\pi/2\) rotation \(z\mapsto iz\), so that the four single monopoles separate on the vertices of the dual tetrahedron. The reverse of Figure \ref{fig:4t}, followed by the flow with the opposite sign for the perturbation parameter, matches a geodesic in \(M_4\) discovered by deriving the associated family of Nahm data \cite{HS1}. The type of initial conditions reported \cite{Bach} for numerical simulations of monopole scattering do not allow the study of this scattering process, but as mentioned earlier, the computation of the fields displayed in the final image in 
Figure \ref{fig:4t}, and related projections to BPS monopoles, is a new method to provide the required initial conditions.

\begin{figure}[!ht]\begin{center}
    \includegraphics[width=1.0\columnwidth]{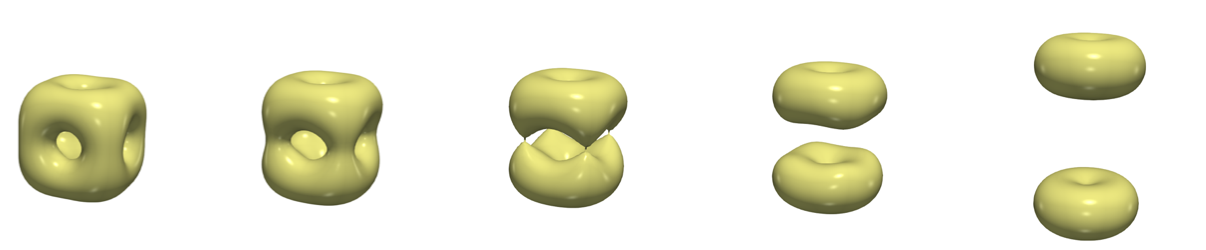}
    \caption{Energy density isosurfaces at increasing times for the cubic charge four monopole with a \(D_{4h}\) symmetric perturbation given by (\ref{map4d4h}) with \(\varepsilon=0.01.\)}
    \label{fig:4d4ha}\end{center}\end{figure}

\begin{figure}[!ht]\begin{center}
    \includegraphics[width=1.0\columnwidth]{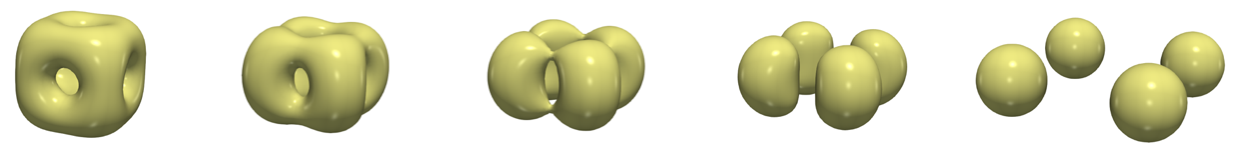}
    \caption{Energy density isosurfaces at increasing times for the cubic charge four monopole with a \(D_{4h}\) symmetric perturbation given by (\ref{map4d4h}) with \(\varepsilon=-0.01.\)}
    \label{fig:4d4hb}\end{center}\end{figure}

The \(O_h\) symmetry of the saddle point solution is broken to \(D_{4h}\) symmetry by breaking the \(C_3\) rotation, but retaining the \(C_2\) rotation \(R(1/z)=R(z),\) via the perturbation
\be
R(z)=\frac{z^4+2i\sqrt{3}(1+\varepsilon)z^2+1}{z^4-2i\sqrt{3}(1+\varepsilon)z^2+1}.
\label{map4d4h}
 \ee
 Figure \ref{fig:4d4ha} reveals that applying this perturbation with \(\varepsilon=0.01\) generates a fission of the cube into a pair of charge two monopoles. Applying the perturbation with \(\varepsilon=-0.01\) produces a fission into four single monopole on the vertices of an expanding square, as displayed in Figure \ref{fig:4d4hb}. The reverse of Figure \ref{fig:4d4ha} followed by Figure \ref{fig:4d4hb} resembles a geodesic in \(M_4\) describing a scattering that instantaneously forms the cube \cite{HMM}. Approximate Nahm data has been used to produce the corresponding energy density isosurface plots \cite{Su5}, and both these are in good agreement with numerical simulations of the same monopole scattering process \cite{Bach}.
 
\begin{figure}[!ht]\begin{center}
    \includegraphics[width=1.0\columnwidth]{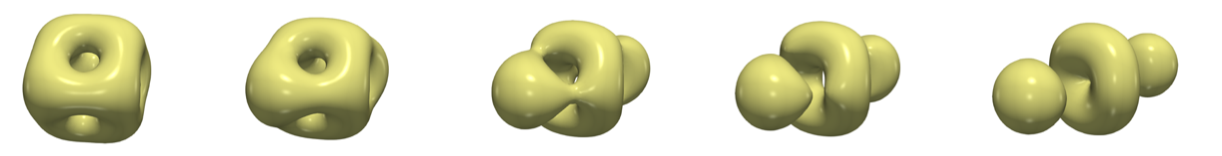}
    \caption{Energy density isosurfaces at increasing times for the cubic charge four monopole with a \(D_{3d}\) symmetric perturbation given by (\ref{map4d3d}) with \(\varepsilon=0.1.\)}
    \label{fig:4d3db}\end{center}\end{figure}

  In an alternative orientation, the cubic map (\ref{map4}) may be written as
 \be
 R(z)=\frac{2\sqrt{2}z^3+1}{z^4-2\sqrt{2}z},
 \ee
 so that the \(C_3\) rotation is more obvious, as it acts as \(R(e^{2\pi i/3}z)=e^{-2\pi i/3}R(z).\) The \(O_h\) symmetry is broken to \(D_{3d}\) symmetry by preserving this \(C_3\) rotation together with the \(C_2\) rotation
 \(R(-1/z)=1/R(z)\) and the reflection \(R(\bar z)=\overline{R(z)}\), via the perturbed map
 \be
 R(z)=\frac{2\sqrt{2}(1+\varepsilon)z^3+1}{z^4-2\sqrt{2}(1+\varepsilon)z}.
 \label{map4d3d}
 \ee
 The resulting gradient flow is presented in Figure \ref{fig:4d3db} for \(\varepsilon=0.1\), where a pair of single monopoles fission from the cube to leave a charge two monopole. Applying the same perturbation with 
\(\varepsilon<0\) does not produce fission, as it flows to the axially symmetric charge four monopole. This can again be understood as a flow between two saddle points, from \(\varepsilon=0\) to \(\varepsilon=-1\), as the map (\ref{map4d3d}) evaluated at \(\varepsilon=-1\) is the axially symmetric map \(R(z)=1/z^4.\) Reversing the sequence in Figure \ref{fig:4d3db} and adjoining the flow from the cube to the axial charge four monopole matches  another example of twisted line scattering \cite{HS3}, with the second half of the associated geodesic in \(M_4\) being a reversal of the first half with the addition of a rotation along the line of approach, so that the cube reforms in a different orientation, with the final outcome being two single monopoles moving away from a charge two monopole along the same line as the initial approach.
 
\subsection{Dodecahedral charge 7}
The \({\cal I}\) minimizing map of degree seven has dodecahedral symmetry and is given by 
\be
R(z)=\frac{7z^6-7\sqrt{5}z^4-7z^2-\sqrt{5}}{\sqrt{5}z^7+7z^5+7\sqrt{5}z^3-7z}.
\label{map7}
\ee
The \(C_5\) symmetry is realized as
\be
R\bigg(\frac{\tau(i\tau-1)z+1}{z+\tau(i\tau+1)}\bigg)=
\frac{\tau^{-1}(i\tau^{-1}+1)R(z)+1}{R(z)+\tau^{-1}(i\tau^{-1}-1)},
\ee
where \(\tau=(1+\sqrt{5})/2\) is the golden ratio. Taken together with the
\(C_3\) rotation given by
\(R((iz+1)/(-iz+1))=(iR(z)+1)/(-iR(z)+1)\), and the reflection \(R(\bar z)=\overline{R(z)},\) this generates 
the dodecahedral symmetry \(Y_h.\)  

\begin{figure}[!ht]\begin{center}
    \includegraphics[width=1.0\columnwidth]{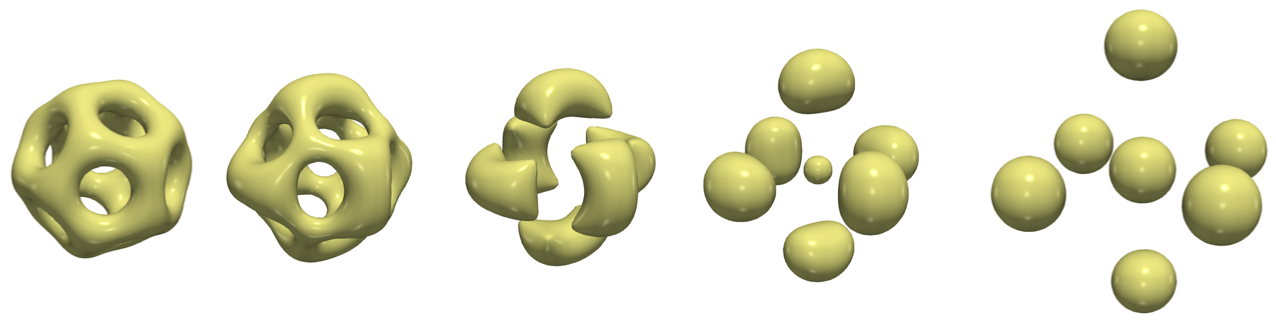}
    \caption{Energy density isosurfaces at increasing times for the dodecahedral charge seven monopole with a \(T_{h}\) symmetric perturbation given by (\ref{map7t}) with \(\varepsilon=0.01.\)}
    \label{fig:7ta}\end{center}\end{figure}

The \(Y_h\) symmetry is broken to tetrahedral \(T_h\) symmetry by breaking the \(C_5\) rotation but retaining the \(C_3\) rotation and the \(C_2\) rotation \(R(-z)=-R(z)\), together with the reflection. The perturbed rational map is 
\be
R(z)=\frac{7(1+\varepsilon)z^6-7\sqrt{5}z^4-7(1+\varepsilon)z^2-\sqrt{5}}{\sqrt{5}z^7+7(1+\varepsilon)z^5+7\sqrt{5}z^3-7(1+\varepsilon)z},
\label{map7t}
\ee
with the evolution resulting from this perturbation for \(\varepsilon=0.01\) displayed in Figure \ref{fig:7ta}, revealing that the dodecahedron fissions into six single monopoles moving along the Cartesian axes, leaving behind a single monopole at the origin.

The tetrahedral map (\ref{map7t}) evaluated at \(\varepsilon=-1\) has an additional symmetry, given by \(R(iz)=iR(z)\), enhancing the tetrahedral to cubic symmetry 
\(O_h.\) Gradient flow using the perturbed map (\ref{map7t}) with \(\varepsilon<0\) flows to the cubic charge seven monopole, this being another example of a flow between saddle points, from \(\varepsilon=0\) to \(\varepsilon=-1\), rather than monopole fission.  
Reversing the sequence in Figure \ref{fig:7ta} and following it by the flow between the dodecahedron and the cube, resembles half a geodesic in \(M_7\), first identified via rational maps \cite{HMS}. The second half of the geodesic is a reversal of the first half with a rotation of \(\pi/2\), so that the dodecahedron is obtained with a different orientation but the six monopoles separate along the Cartesian axes, making the incoming and outgoing lines of approach the same. The Nahm data is not known for this geodesic, so energy density isosurfaces are not available for this family of BPS monopoles. However, a related family of instantons has been obtained 
and used to create Skyrme fields \cite{SiSu}, via the Atiyah-Manton instanton holonomy construction \cite{AM,AM2}, with results that closely resemble those in Figure \ref{fig:7ta}.

\begin{figure}[!ht]\begin{center}
    \includegraphics[width=1.0\columnwidth]{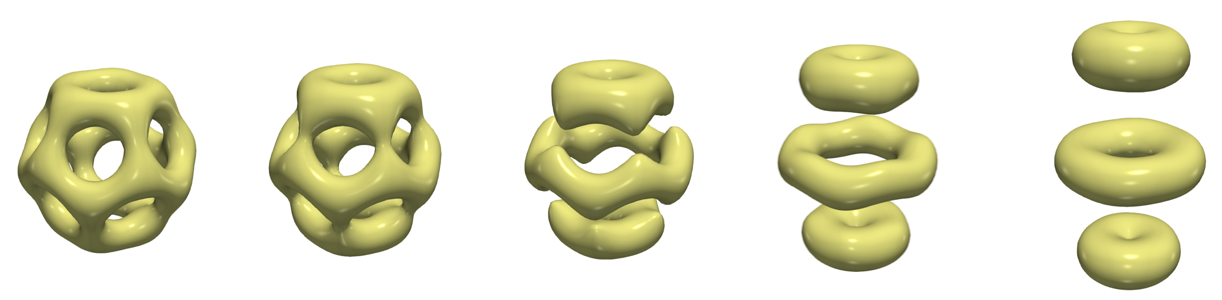}
    \caption{Energy density isosurfaces at increasing times for the dodecahedral charge seven monopole with a \(D_{5d}\) symmetric perturbation given by (\ref{map7d5d}) with \(\varepsilon=0.1.\)}
    \label{fig:7d5da}\end{center}\end{figure}

In an alternative orientation, the dodecahedral map (\ref{map7}) may be written as
\be
R(z)=\frac{z^7-7z^5-7z^2-1}{z^7+7z^5-7z^2+1},
\ee
so that the
\(C_5\) rotation acts as  
\be
R(e^{2\pi i/5}z)=\frac{(e^{4\pi i/5}+1)R(z)+e^{4\pi i/5}-1}
{(e^{4\pi i/5}-1)R(z)+e^{4\pi i/5}+1}.
\ee
The \(Y_h\) symmetry is broken to dihedral \(D_{5d}\) symmetry by preserving this \(C_5\) rotation, the \(C_2\) rotation given by \(R(-1/z)=-1/R(z)\), and the reflection \(R(\bar z)=\overline{R(z)}\). The required perturbed rational map is
\be
R(z)=\frac{z^7-7(1+\varepsilon)z^5-7(1+\varepsilon)z^2-1}{z^7+7(1+\varepsilon)z^5-7(1+\varepsilon)z^2+1},
\label{map7d5d}
\ee
and Figure \ref{fig:7d5da} presents the gradient flow evolution using this map with \(\varepsilon=0.1.\) The fission proceeds via the top and bottom faces of the dodecahedron pulling apart, so that each forms a charge two monopole, leaving behind a charge three monopole that is approaching the axially symmetric solution. 

The flow from 
(\ref{map7d5d}) with \(\varepsilon<0\) is another flow between saddle points, as expected because the rational map 
(\ref{map7d5d}) evaluated at \(\varepsilon=-1\) is an \(SU(2)\) M\"obius transformation of the axially symmetric degree seven map \(R(z)=z^7.\) The reverse of Figure \ref{fig:7d5da} followed by the flow from the dodecahedron to the axially symmetric charge seven monopole, is another match for half a geodesic of the twisted line scattering type \cite{HS3}. The second half of this geodesic in \(M_7\)
is a reversal of the first half plus a rotation by \(\pi/5\) along the common axis of the three separating monopoles, of charges two, three and two, all of which become axially symmetric in the limit that the charge two monopoles tend to infinity.

\begin{figure}[!ht]\begin{center}
    \includegraphics[width=1.0\columnwidth]{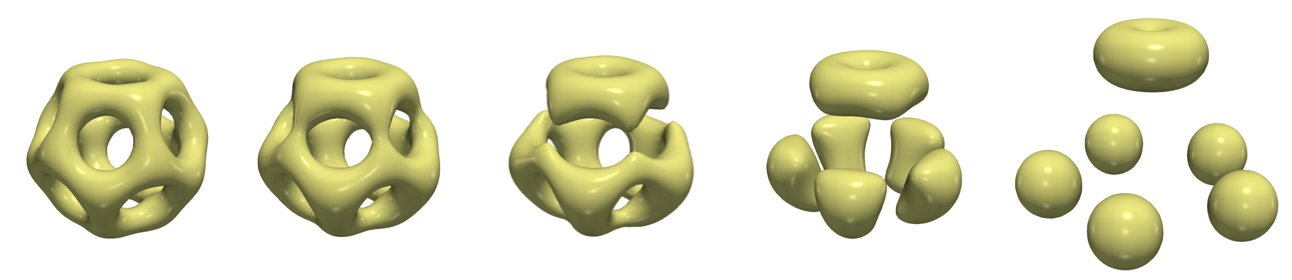}
    \caption{Energy density isosurfaces at increasing times for the dodecahedral charge seven monopole with a \(C_{5v}\) symmetric perturbation given by (\ref{map7c5}) with \(\varepsilon=0.01.\)}
    \label{fig:7c5a}\end{center}\end{figure}

The dodecahedral symmetry can be broken further to cyclic  \(C_{5v}\) symmetry by breaking the \(C_2\) rotation using
the perturbed rational map
\be
R(z)=\frac{z^7-7(1-\varepsilon)z^5-7(1+\varepsilon)z^2-1}{z^7+7(1-\varepsilon)z^5-7(1+\varepsilon)z^2+1}.
\label{map7c5}
\ee
The evolution from this perturbation with \(\varepsilon=0.01\) is displayed in Figure \ref{fig:7c5a}, which reveals the fission into a charge two monopole moving away from a plane containing five single monopoles on the vertices of an expanding pentagon.

\subsection{Buckyball charge 17}  
\begin{figure}[!ht]\begin{center}
    \includegraphics[width=1.0\columnwidth]{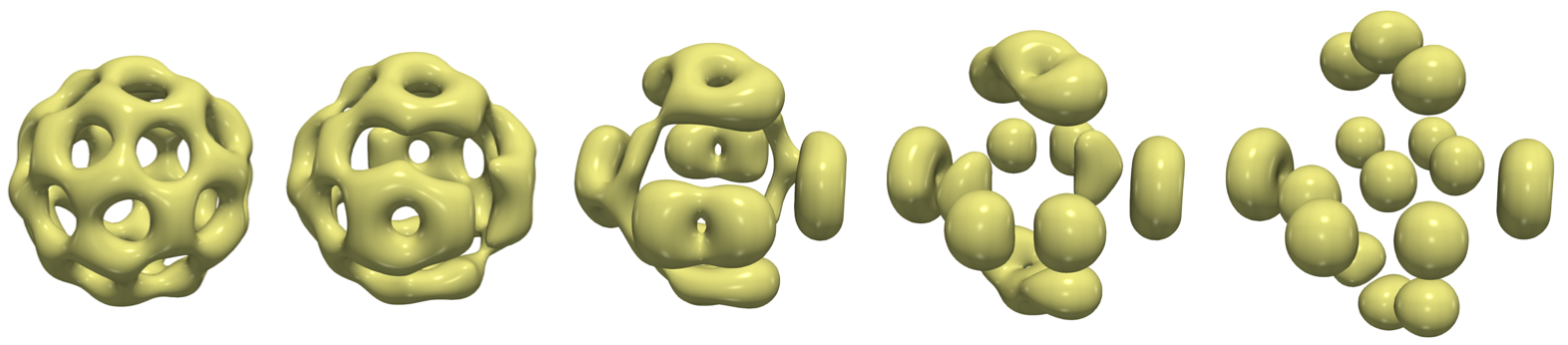}
    \caption{Energy density isosurfaces at increasing times for the buckyball charge seventeen monopole perturbed by the cubic boundary of the simulation grid.}
    \label{fig:17yb}\end{center}\end{figure}

If the size of the monopole is not much smaller than the simulation region, then the cubic boundary of the numerical grid can introduce an artificial perturbation that seeds monopole fission of a saddle point solution. As an example, consider the charge seventeen monopole with dodecahedral symmetry \(Y_h.\) The \({\cal I}\) minimizing map of degree 
seventeen is given by 
\be
R(z)=\frac{17z^{15}-187z^{10}+119z^5-1}{z^{17}+119z^{12}+187z^7+17z^2},
\label{map17}
\ee
and the associated monopole has its energy density localized on the edges of a truncated icosahedron (see the first image in Figure \ref{fig:17yb}), familiar as
the buckyball in carbon chemistry. The cubic boundary breaks the \(Y_h\) symmetry to the cyclic symmetry \(C_{2h}\), generated by \(R(-1/z)=-1/R(z)\) and \(R(\bar z)=\overline{R(z)}.\)
The gradient flow evolution starting from the map (\ref{map17}) is shown in Figure \ref{fig:17yb}, where it can be seen that the buckyball fissions into thirteen single monopoles and a pair of charge two monopoles. The fission is less symmetric than the controlled examples studied earlier, preserving only the \(C_{2h}\) symmetry, which is not very restrictive for a monopole charge as high as seventeen.
    
\section{Conclusion}\quad
Numerical simulations of the gradient flow equations for \(SU(2)\) Yang-Mills-Higgs theory, with a Higgs potential, have provided a study of the fission of a variety of saddle point monopole solutions. The investigation has been facilitated by the use of rational maps between Riemann spheres, to generate the initial conditions for the monopole solutions and to control the symmetry of the perturbations that seed monopole fission. The results have been compared to known geodesics in the BPS monopole moduli space that describe monopole scattering, demonstrating a qualitative agreement between families of monopole configurations.

A potential application of the approach presented in this paper is to provide initial monopole fields for numerical simulations of monopole scattering, including in the BPS limit, as the flow can be switched to project to points in the BPS monopole moduli space. This could be a helpful approach to providing initial conditions, given that there is currently no generally applicable method to create a suitable arbitrary superposition of monopoles with prescribed internal phases. Current approaches are limited to bespoke cyclic arrangements with phases that yield planar scattering, or a pair of axially symmetric monopoles that share the same symmetry axis \cite{Bach}. 

It might be interesting to attempt to map out the energy landscape of non-BPS monopole configurations, at least for some families that include multiple saddle points, such as the twisted line examples. The methods presented in this paper are appropriate for this task, although larger and finer numerical grids are required to obtain the energies to an appropriate accuracy to capture small energy differences between varied configurations in the landscape. The algebraic decay of the gauge potential, in contrast to the exponential decay of the Higgs field, is a challenge in obtaining very accurate energy computations in three-dimensional simulations. 

The present study is restricted to \(SU(2)\) monopoles, but the correspondence between BPS monopoles and rational maps from the Riemann sphere extends to any compact semisimple gauge group, with the target space of the rational map given by a flag manifold appropriate to the type of symmetry breaking \cite{Jar}. It should therefore be possible to extend the approximation for monopole fields to gauge groups beyond \(SU(2)\), to provide initial conditions to compute saddle points and investigate monopole fission. In the related case of Skyrmions, the rational map approximation has been generalized from \(SU(2)\) to other groups \cite{IPZ}. 

Finally, an avenue for future research is to extend the investigation from monopoles in Euclidean space to monopoles in hyperbolic space. In hyperbolic space the BPS limit again produces a moduli space of static monopole solutions with arbitrary positions and phases, with an analogous correspondence between hyperbolic BPS monopoles and rational maps between Riemann spheres \cite{JN}. However, the geodesic approximation to monopole scattering no longer applies in hyperbolic space, as the natural metric on moduli space is divergent. As a consequence, very little is known about the dynamics of hyperbolic monopoles, and there have been no numerical simulations of their scattering. Studying the gradient flow of non-BPS hyperbolic monopoles could be a first step in investigating any type of hyperbolic monopole dynamics.

\end{document}